\documentclass[%
 reprint,
superscriptaddress,
 amsmath,amssymb,
 prl,
]{revtex4-1}

\usepackage{graphicx}
\usepackage{dcolumn}
\usepackage{bm}
\usepackage{braket}
\usepackage[title]{appendix}
\usepackage{xcolor}
\usepackage{hyperref}
\hypersetup{colorlinks,
citecolor = blue,
urlcolor = blue}

\begin{document}

\preprint{APS/123-QED}

\title{Exciton migration in two-dimensional materials}

\author{Mikhail Malakhov}%
\affiliation{Departamento de Qu\'{i}mica, Universidad Aut\'{o}noma de Madrid, 28049 Madrid, Spain}
\affiliation{Instituto de Ciencia de Materiales de Madrid (ICMM), Consejo Superior de Investigaciones Científicas (CSIC), Sor Juana Inés de la Cruz 3, 28049 Madrid, Spain}
\author{Giovanni Cistaro}
\affiliation{Theory and Simulation of Materials (THEOS), École Polytechnique Fédérale de Lausanne (EPFL), CH-1015
Lausanne, Switzerland}

\author{Fernando Mart\'in}
\affiliation{Departamento de Qu\'{i}mica, Universidad Aut\'{o}noma de Madrid, 28049 Madrid, Spain}
\affiliation{Instituto Madrile\~no de Estudios Avanzados en Nanociencia (IMDEA-Nanociencia), Cantoblanco, 28049 Madrid, Spain}

\author{Antonio Pic\'on}
\email{antonio.picon@uam.es, corresponding author}
\affiliation{Departamento de Qu\'{i}mica, Universidad Aut\'{o}noma de Madrid, 28049 Madrid, Spain}

\date{\today}

\begin{abstract}
Excitons play an essential role in the optical response of two-dimensional materials. These are bound states showing up in the band gaps of many-body systems and are conceived as quasiparticles formed by an electron and a hole. By performing real-time simulations in hBN, we show that an ultrashort (few-fs) UV pulse can produce a coherent superposition of excitonic states that induces an oscillatory motion of electrons and holes between different valleys in reciprocal space, leading to a sizeable exciton migration in real space. We also show that an ultrafast
spectroscopy scheme based on the absorption of an attosecond pulse in combination with the UV pulse can be used to read out the laser-induced coherences, hence to extract the characteristic time for exciton migration. This work opens the door towards ultrafast electronics and valleytronics adding time as a control knob and exploiting electron coherence at the early times of excitation. 

\end{abstract}

\maketitle




Electrons usually move much faster than nuclei and, for this reason, they play a dominant role in the optical response of two- and three-dimensional materials. Thus, manipulating and controlling electronic motion in its natural timescale, well before the lattice has time to respond, may open an unprecedented platform for charge transport and valleytronics based on electron coherence. Nowadays, we have the technology to produce laser pulses as short as several attoseconds (10$^{-18}$ s), which enables to track and investigate electron dynamics \cite{Geneaux_HHF_review_2019, Zong2023_review}. By using such technology, techniques such as ultrafast absorption spectroscopy have been carried out to observe electron motion in insulators, semiconductors, and semimetals \cite{Geneaux_HHF_review_2019, Zong2023_review, Schultze_silicon_2014, Zurch_germanium_2017, Lucchini_diamond_2016, Schlaepfer_GaAs_2018, Volkov_transition_metals_2019}, and even in few-layers materials \cite{Buades_XANES_2021}.

Attosecond and few-femtosecond pulses not only enable to track electron dynamics, but also to trigger unusual charge dynamics. In particular, real-time investigations of charge migration in molecular systems, induced  by such ultrashort pulses, have been systematically reported in the literature for almost a decade, providing an unprecedented understanding of the process and opening the way for new control schemes of chemical reactions (the so-called attochemistry) \cite{Breidbach_2005, Remacle_chemistry_2006, Calegari_phenylalanine_2014, Nisoli_AED_Molec_2017}. Charge migration can be induced by  creating a coherent superposition of {\it bound} molecular states with a broadband, attosecond, or few-fs pulse. During the free propagation of the system, this coherent superposition induces fast oscillations between the involved states, which, in the case of covering different spatial regions and/or exhibiting quite different electronic properties, may translate into electron transfer from one side of the molecule to another. To our knowledge, similar charge migration processes have not yet been observed in condensed-matter systems, probably because the electrons organize in a quasi continuum of delocalized states (the electronic bands), so that any coherent superposition induced by an ultrashort pulse involves a huge number of states with continuously and smoothly varying electronic properties.

\begin{figure*}[t]
\includegraphics[width=0.99\textwidth]{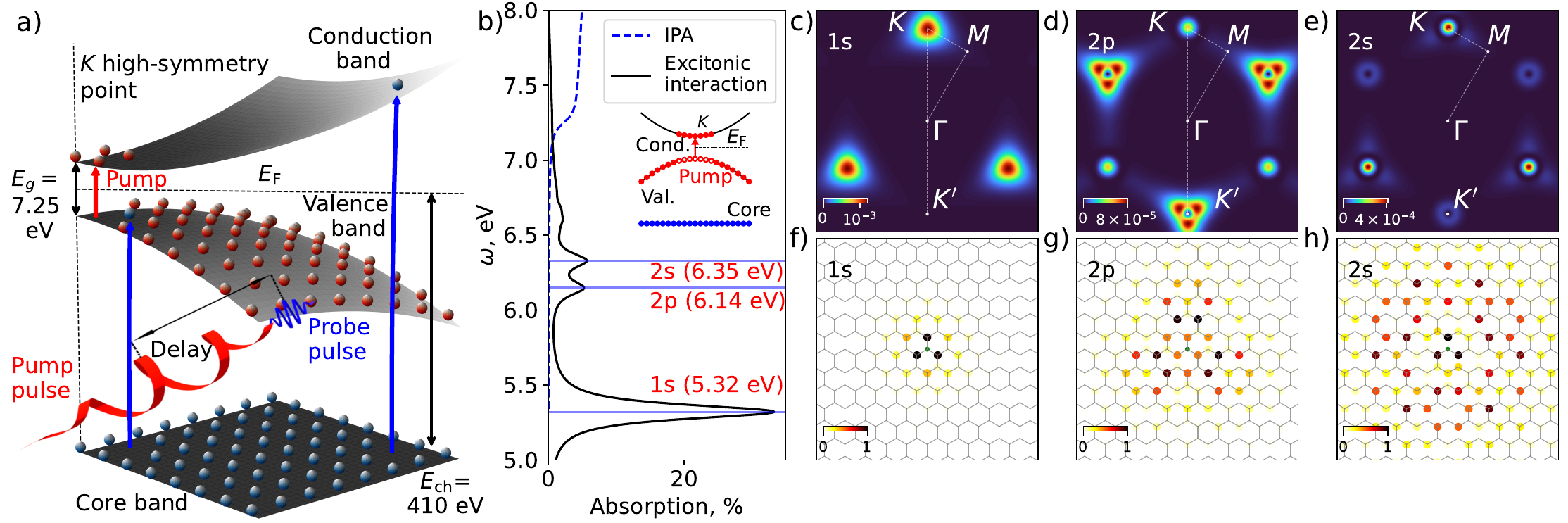}
  \caption{ {\bf Two-color excitations in hBN}. a) Illustration of the ultrafast scheme and the possible transitions in hBN.  b) The UV absorption spectrum resulting from our real-time simulations using the EDUS code \cite{EDUS_JCTC_2023}. Dashed blue line represents the independent particle approximation (IPA) calculations of the absorption when no electron-electron interactions are included. The peaks correspond to the 1s, 2p, and 2s excitons at 5.32, 6.14, and 6.35 eV, respectively.  c)-e)  Distribution in ${\bf k}$ space for the three excitons, obtained by using a long 120-fs pulse resonant to the corresponding exciton peak and circularly polarized light. f)-h) The real-space distribution of the three excitons.
  }\label{fig:SpectrumUV}
\end{figure*}

In this manuscript, we demonstrate that laser induced charge migration is possible in materials whose optical response is dominated by excitonic interactions. Excitons can be considered as quasi-particles composed of an electron-hole pair bound via Coulomb interaction. This interaction can only manifest in systems where screening effects are not dominant, as, e.g.,   non-metallic two-dimensional (2D) materials, where mobility of the remaining electrons is hampered due to the reduced dimensionality. As a consequence, the optical response of non-metallic 2D materials is almost entirely dominated by excitons \cite{Wang_ColloquiumExcitons_2018}.  Excitons are usually associated to bound states located within bandgaps.
The exciton migration can thus be induced when an ultrashort pulse excites a superposition of those quasi-particle states. In this work, we have performed real-time simulations for monolayer boron nitride (hBN) interacting with an ultrashort UV pulse using our recently developed EDUS approach \cite{EDUS_JCTC_2023}. hBN displays two valley pseudospins \cite{Xiao_Valley_2007, Xiao_Berry_2010} related to its inversion symmetry and electronic structure  \cite{Yao_Valley_2008}. We show that the ultrashort pulse enables us to excite a superposition of s- and p- excitons that are localized in different valleys of the reciprocal space (and also different regions in real space). The exciton migration produces the oscillation of excitons from one valley, around the K point, to the other valley, around the K' point, in about 10 fs (10$^{-15}$ s). This oscillation is translated into fast beatings in the laser-induced current. Finally, we show that such fast oscillations can be read out by using X-ray attosecond transient absorption spectroscopy (ATAS)\cite{Geneaux_HHF_review_2019}, which has been successfully used to study other electron dynamics processes in insulators \cite{Moulet_core_exciton_2017,Geneaux_core_exciton_2020,Lucchini_core_exciton_2021}.



\begin{figure}[t]
\includegraphics[width=0.49\textwidth]{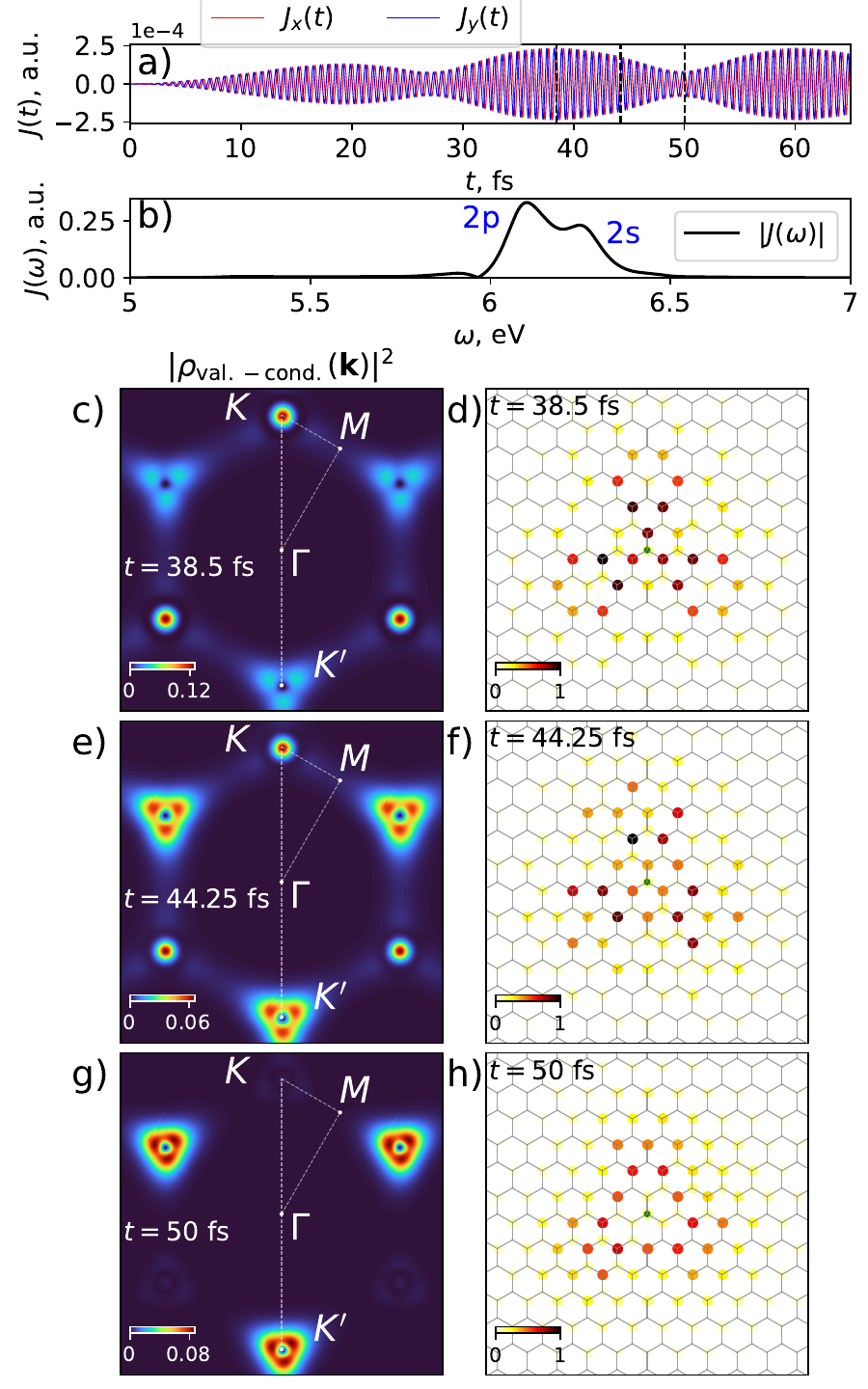}
  \caption{ {\bf Exciton migration.} a) Time evolution of the current induced by the circularly polarized UV pulse. b) Fourier transform of the laser-induced current. Figs c)-h) show snapshots, at the times indicated in a), of the excitonic distributions in reciprocal and real space.
  }\label{fig:Time_evolution}
\end{figure}

Numerical simulations of ATAS including excitonic effects are challenging. First, because ultrafast schemes require a theory beyond linear response, as one has to describe the absorption of at least two photons (one from the pump pulse, another one from the probe pulse). Second, because valence excitons as those considered in this work can only be described by correctly accounting for the electron-electron interactions. In this respect, some significant progress has been performed at the level of real-time TDDFT  \cite{Sun_TDDFT_2021} and real-time Green's function based methods \cite{Perfetto_RealTimeGW_2022}, and in our numerical implementation of the semiconductor Bloch equations; EDUS \cite{EDUS_JCTC_2023}. In addition, EDUS reduces the computational cost of including core orbitals and enables to describe x-ray interactions. In previous work \cite{EDUS_JCTC_2023}, we have shown, by explicit comparison with elaborate calculations for single-photon absorption \cite{Galvani_Excitons_2016, Ridolfi_Excitons_2020, uria_excitonshBN2023}, that a reasonable description of the 2p and 2s excitons of hBN can be achieved even at the tight-binding level using a two-band model. Thus, to face the more challenging ATAS scenario, here we have followed a similar approach and used a three-band (two valence + one K-shell) tight-binding model of hBN, see the illustration in Fig. \ref{fig:SpectrumUV}a. The core band is flat and belongs to the 1s orbital of N, at an energy of $E_{ch}=410$ eV with respect to the Fermi level. In brief, we solve the real-time electron dynamics with the EDUS code \cite{EDUS_JCTC_2023,CistaroGraphene_2021,Picon_2019_X_Ray}, which consists in evolving the one-electron reduced density matrix in the reciprocal space, see more details in SI. The laser-matter and electron-electron interactions that give rise to excitonic effects are accounted for on an equal footing in the time domain. Electron-electron interactions are taken into account in the dynamical mean-field approximation, which is a reasonable approximation to describe excitons \cite{KIRA2006155}. Via the calculated one-electron density matrix in time, we are able then to obtain the polarization of the system and thence the absorption spectrum. 

In our simulations, an 11.3-fs FWHM pulse centered at a photon energy of 6.14 eV and circularly polarized, depicted in Fig. \ref{fig:SpectrumUV}a, interacts with hBN.
Several exciton peaks are present in the UV spectrum within the bandgap of $E_g=7.25$ eV, see Fig. \ref{fig:SpectrumUV}b. When  electron-electron (excitonic) interactions are switched off, the absorption only takes place for photon energies above the energy bandgap $E_g$. When excitonic interactions are included, strong absorption peaks appear within the bandgap. The prominent peaks have s- and p- characters, which present a distinctive distribution in {\bf k} space, see Figs. \ref{fig:SpectrumUV}c-e in which the off-diagonal part of the density matrix is represented. The 1s exciton is well-localized around the K points. Note that an opposite handness of the polarization would excite the degenerate 1s exciton that is located at the valley around the K' point \cite{Yao_Valley_2008}. The next peak corresponds to the 2p exciton, which is also degenerate. If we use the same handness for the polarization, then we excite the 2p exciton, which is mainly localized and shows a clear singularity around the K' points. This exciton is hybridized with the 2s exciton, which is key to induce the exciton migration, and shows a small population around the K points. The third peak corresponds to the 2s exciton, which is also degenerate, and our chosen circular polarization excites the K valleys. Because of the hybridization, this exciton also has a 2p component in the K' valleys. We represent the three excitons in real space in Figs. \ref{fig:SpectrumUV}f-h. Note that the 2s exciton is well-localized in {\bf k} space and quite spread in real space.

\begin{figure}[t]
\includegraphics[width=0.49\textwidth]{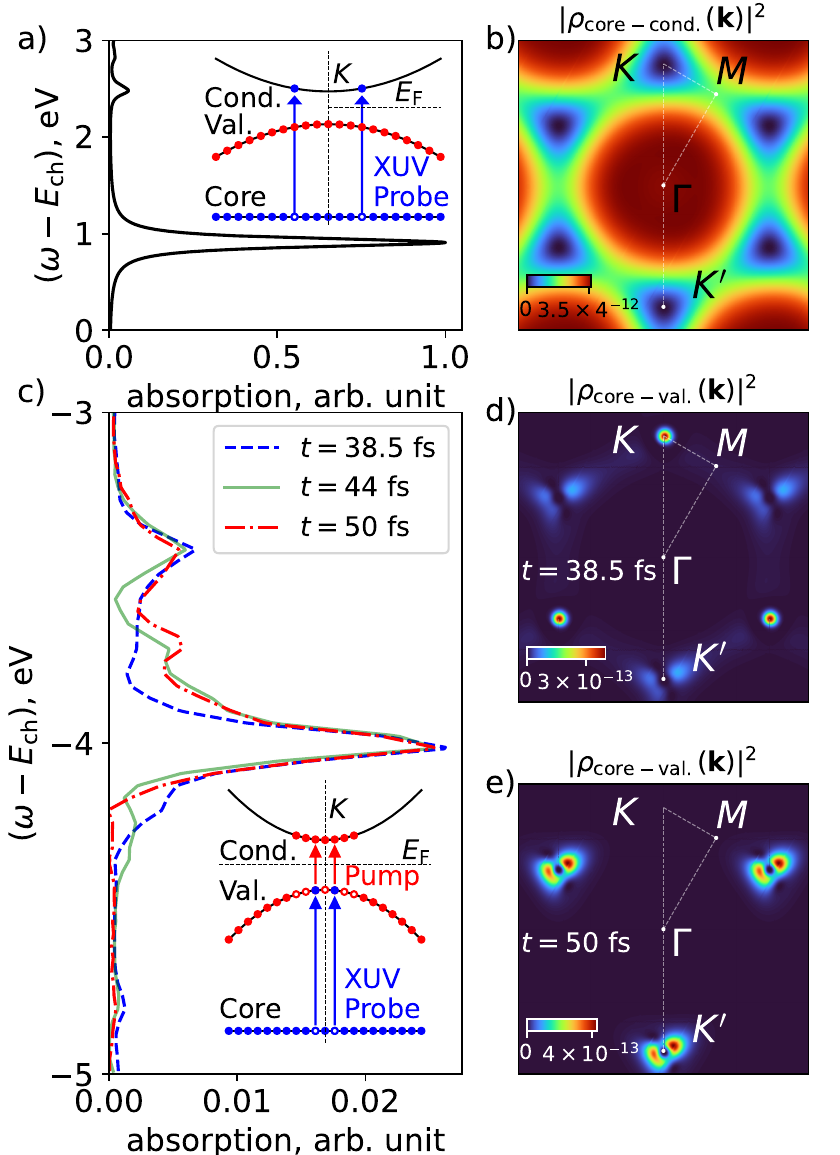}
  \caption{{\bf Laser-induced core excitons.} a) N K-edge absorption spectrum of the attosecond x-ray pulse in the absence of the pump pulse. Note we take as a reference the energy excitation from the 1s N orbital to the Fermi level. b) Distribution of the population in {\bf k} space with no pump excitation. c) N K-edge absorption spectrum of the attosecond x-ray pulse after pump excitation, the time delay corresponding to the times given in Fig. \ref{fig:Time_evolution}.  d)-e) Distribution in {\bf k} space for the square modulus of the off-diagonal density matrix of the core-valence part.
  }\label{fig:Core_exciton}
\end{figure}

\begin{figure*}[t]
\includegraphics[width=0.99\textwidth]{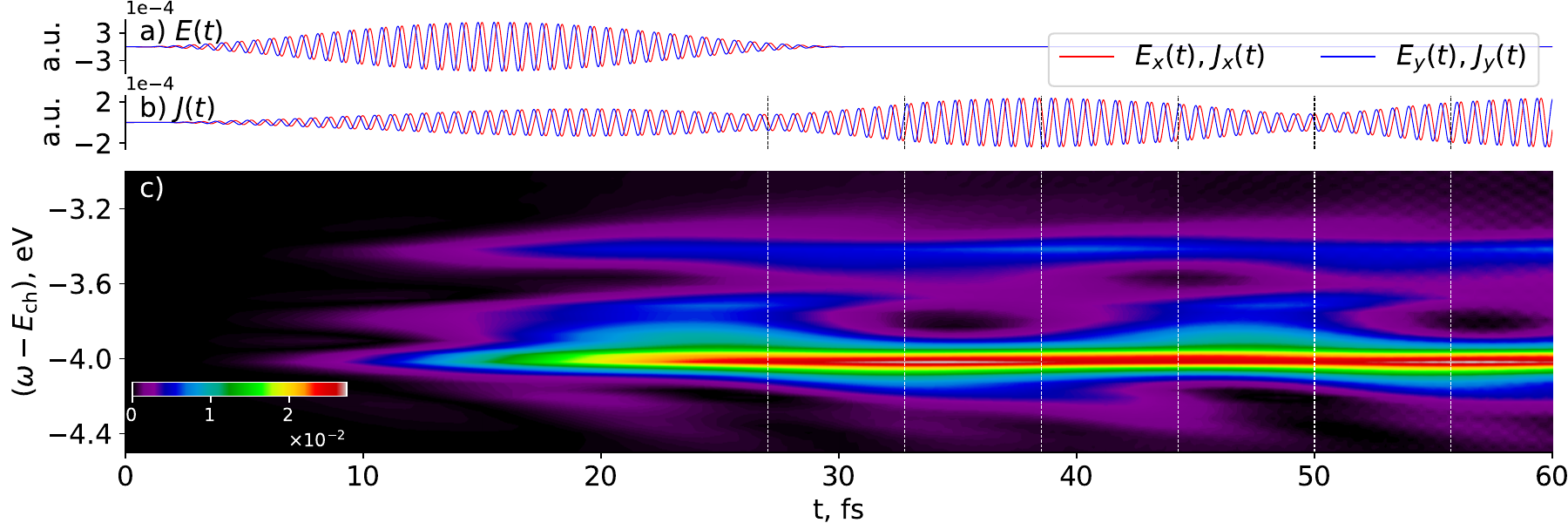}
  \caption{ {\bf Attosecond transient absorption spectroscopy at the valence band for tracking exciton migration.} a) Pump pulse in time, b) Laser-induced current, and c) ATAS features at the valence band. Dashed vertical lines indicate maxima and minima, and intermediate points, of quantum beats in current.
  }\label{fig:ATAS}
\end{figure*}

The ultrashort pulse is broad enough in frequency to excite both the 2p and 2s excitons. The laser-induced current in time, see Fig. \ref{fig:Time_evolution}a, presents some quantum beats due to the coherent excitation of the two excitonic states, see the Fourier transform of the current in Fig. \ref{fig:Time_evolution}b. Those quantum beats will last until the coherence is lost. Electron collisions or electron–phonon couplings may contribute to the dephasing, but here we neglect these effects, which will have a minor impact in such short time scale. If we plot the time evolution of the density matrix in {\bf k} space in a maximum of the beating, see Fig. \ref{fig:Time_evolution}c, and in a minimum of the beating, see Fig. \ref{fig:Time_evolution}g, we observe that it mainly moves from the K to the K' valley in approximately 10 fs. Then it goes back to its initial state in the next 10 fs.  Hence, at some points in time (e.g., 38.5 fs) the exciton is well-localized at the K' valley with s-character, while at other times (e.g., 50 fs) is mainly localized at the K valley with an p-character. The period of the oscillation $\tau\sim20$ fs is linked to the energy difference $\Delta E$ of the exciton states through the formula $\Delta E = 2\pi / \tau$. Therefore, the exciton excitation and energies may be a control knob to tailor the migration oscillations in time. In real space, see Figs. \ref{fig:Time_evolution}d,f,h, we observe how the spatial distribution significantly changes when the exciton is localized around the K valleys and moves to the K' valleys. The real-space distribution is mainly a linear superposition of the 2s and 2p distribution given in Figs. \ref{fig:SpectrumUV}d-e. For example, one observes that at $t=38.5$ fs the distribution clearly shows the structure of the 2p, with strong population in the three first neighbours as in the 2s structure. This real-space motion is connected to the calculated beats in the current.

These results show that coherent population of excitonic states with different degrees of delocalization in the crystal may induce a fast electron/hole motion that controls the current and the valley population in a short time scale. This phenomenon relies on the coherence of the exciton states and therefore it should be observed before the laser-induced coherent dynamics couples to other degrees of freedom. To experimentally observe exciton migration, here we propose to use ATAS. 
ATAS consists in sending a second laser pulse, the so-called probe pulse, to the system with a certain time delay with respect to the pump pulse, see Fig. \ref{fig:SpectrumUV}a, and then in measuring the absorption of the probe pulse as a function of the time delay. In this particular case, we consider an attosecond pulse that excites the K-edge transitions at the nitrogen site, i.e. transitions that promote electrons from the 1s orbitals of nitrogen to the valence/conduction band. In our tight-binging model we include this additional core band, see more details in the SI. The attosecond pulse has a 133-as FWHM duration, see SI, and a photon energy centered around 410 eV. The bandwidth of the pulse is large enough to cover all transitions to the valence and conduction bands. With no pump pulse, the calculated absorption shows a prominent core-exciton peak above the Fermi level, see Fig. \ref{fig:Core_exciton}a. This peak arises from the conduction band, as core electrons cannot be promoted into the fully occupied valence band. The transfer of core electrons to the conduction band is clearly illustrated by the relative population of core and conduction bands shown in Fig. \ref{fig:Core_exciton}b. As can be seen, besides the areas around the K,K' points, the whole reciprocal space is partially populated. This is because the conduction band at those areas is dominated by the 2p orbitals of boron and we are exciting from the 1s orbital of nitrogen, which is very well-localized in space.

When the UV pulse is also present, holes created in the valence band, either through exciton formation or promotion of electrons to the conduction band, can be refilled by N K-shell electrons excited by the attosecond X-ray pulse, leading to distinct peaks appearing below the Fermi level -see scheme in the inset of Fig. 3c.
Our real-time simulations show that the shape and magnitude of these peaks are indeed sensitive to exciton migration, see Fig. \ref{fig:Core_exciton}c. The valence band around the K, K' points is dominated by the 2p orbitals of nitrogen. This enhances the transitions from the 1s to the 2p orbitals of nitrogen. Hence, X-ray excitations are very sensitive to any changes occurring around the reciprocal regions in which the main exciton migration takes place. The exciton migration changes the hole distribution in the valence band with time, so that the refilling in the valence band with core electrons will also depend on time, leading to different structures of the core-exciton peaks that are formed. When the 2p character is dominant, the valence hole distribution is more extended in the reciprocal space and this enables access to more core exciton states. This is clearly shown in Fig. \ref{fig:Core_exciton}d-e, where we represent the off-diagonal part of the density matrix between the core and valence band in reciprocal space at two different pump-probe delays. As can be seen, the distributions giving rise to core excitons look substantially different at different times. 

Finally, we show in Fig. \ref{fig:ATAS}
the calculated ATAS during and after the pump excitation. Interestingly, we observe how the population of core excitons significantly increases after the maximum intensity of the pump pulse, following the population of the produced holes in the valence band. Those peaks are around 1\% of the intensity of the main core excitons peaks, see Fig. \ref{fig:Core_exciton}a, and appear in an energy window in which there was no absorption of the probe pulse, enabling a high contrast in an experimental measurement. The peaks at -4.0 and -3.4 eV oscillate with the exciton migration, but the most striking feature is found in the energy windows between those peaks, in which we observe the emergence and disappearance of core-exciton peaks. This energy window is ideal for transient absorption measurements in order to read out the fast exciton dynamics and link the absorption peaks to the valence-hole distribution as discussed above.  


In conclusion, we demonstrate the possibility of inducing charge migration in materials whose optical response is dominated by excitons. Using real-time simulations that account both for light-matter and electron-electron interactions, we show how an ultrashort UV pulse induces a superposition of exciton states in hBN that triggers exciton migration between the K and K' valleys. The migration depends on the exciton energies and laser excitation. Furthermore, we show that this dynamics can be read out by performing X-ray ATAS at the nitrogen K edge, which enables to probe the electron/hole density around individual atomic sites.

2D materials offer an ideal platform for controlling the exciton properties via control of layers, substrates, Van der Waals heterostructures, or strain engineering \cite{Chaves_Bandgap_engineering_2020}. Also, exciton energies can be modified by another pulse via laser-induced Stark shift \cite{Slobodeniuk_valley_excitons_2023}. Thus, 
our study not only advances our understanding of exciton dynamics in attosecond science, but it also opens a promising perspective of exploiting exciton migration for developing transport and valleytronics schemes beyond the strong-field regime \cite{Langer_valleytronics_2018, Yamada_valley_2023, Slobodeniuk_valley_excitons_2023, Kobayashi_Floquet_2023}, harnessing ultrafast electronics in two-dimensional materials at the ultimate time scale.


\section*{Acknowledgments}
This publication is based upon work from COST Action AttoChem, CA18222 supported by COST (European Cooperation in Science and Technology). 
M. Malakhov, G. Cistaro, and A. Pic\'{o}n acknowledge grant ref. PID2021-126560NB-I00 (MCIU/AEI/FEDER, UE), and grants refs. 2017-T1/IND-5432 and 2021-5A/IND-20959 (Comunidad de Madrid through TALENTO program). F. Mart\'{\i}n acknowledges the projects PID2019-105458RB-I00 funded by MCIN/AEI/10.13039/501100011033 and by the European Union "NextGenerationEU"/PRTRMICINN programs, and the "Severo Ochoa" Programme for Centres of Excellence in R\&D (CEX2020-001039-S). Calculations were performed at the Centro de Computaci\'on Cient\'ifica de la Universidad Aut\'onoma de Madrid (FI-2021-1-0032), Instituto de Biocomputación y Física de Sistemas Complejos de la Universidad de Zaragoza (FI-2020-3-0008), and Barcelona Supercomputing Center (FI-2020-1-0005, FI-2021-2-0023, FI-2021-3-0019) and Picasso (FI-2022-1-0031,FI-2022-2-0031,FI-2022-3-0022). 






\bibliography{ATAS_Superposition}

\end{document}